\documentclass[11pt,twocolumn,aps,prl,citecolor=blue,reprint,superscriptaddress]{revtex4-1}

\usepackage{pst-node,graphicx}
\usepackage{braket}
    
\usepackage{graphicx,amsmath,amssymb,braket}
\usepackage[colorlinks=true,linkcolor=blue]{hyperref}
\usepackage{ulem}

 \usepackage[colorlinks=true,linkcolor=blue]{hyperref}
 
\begin{document}

\title{Singularity-enhanced terahertz detection in high-mobility field-effect transistors}

\author{M. Khavronin}
\affiliation{Laboratory of 2d Materials for Optoelectronics, Moscow Institute of Physics and Technology, Dolgoprudny 141700, Russia}

\author{A. Petrov}
\affiliation{Laboratory of 2d Materials for Optoelectronics, Moscow Institute of Physics and Technology, Dolgoprudny 141700, Russia}

\author{A.E. Kazantsev}
\affiliation{Physics Department, M.V. Lomonosov Moscow State University, Moscow, Russia}

\author{E.I. Nikulin}
\affiliation{Physics Department, M.V. Lomonosov Moscow State University, Moscow, Russia}

\author{D.A. Bandurin}
\affiliation{Laboratory of 2d Materials for Optoelectronics, Moscow Institute of Physics and Technology, Dolgoprudny 141700, Russia}
\affiliation{Department of Physics, Massachusetts Institute of Technology, Cambridge, Massachusetts 02139, USA}
\email{bandurin@mit.edu}

\begin{abstract}
Detectors of high-frequency radiation based on high-electron-mobility transistors benefit from low noise, room-temperature operation, and the possibility to perform radiation spectroscopy using gate-tunable plasmon resonance. Despite successful proof-of-concept demonstrations, the responsivity of transistor-based detectors of THz radiation, at present, remains fairly poor. To resolve this problem, we propose a class of devices supporting singular plasmon modes, i.e. modes with strong electric fields near keen electrodes. A large plasmon-enhanced electric field results in amplified non-linearities, and thus efficient ac-to-dc conversion. We analyse sub-terahertz detectors based on a two-dimensional electron system (2DES) in the Corbino geometry as a prototypical and exactly solvable model, and show that the responsivity scales as $1/r_0^{2}$ with the radius of the inner contact $r_0$. This enables responsivities exceeding $10$ kV/W at sub-THz frequencies for nanometer-scale contacts readily accessible by modern nanofabrication techniques.

\begin{center}
	\textbf{*Email:} bandurin.d@gmail.com 
\end{center}
\end{abstract}

\maketitle

Excitation of plasmons in radiation detectors provides a convenient tool to enhance absorption and thereby increase sensitivity~\cite{Plasmonic_detector_1,Plasmonic_detector_2,Plasmonic_detector_3}. Such  enhancement in the visible and infrared ranges is easily achieved by the integration of metal nano-objects and photo-sensitive semiconductors~\cite{Plasmonic_detector_nanoparticles}. This technique, however, cannot be extended to terahertz (THz) frequencies due to the high density of electrons in metals. This can, however, be achieved with plasmons bound to the semiconductors themselves~\cite{Knap_resonant,Kukushkin_resonant,Peralta_resonant}. Apart from low carrier density, observation of plasmon resonance in the THz range requires long electron momentum relaxation time $\tau_p$, such that quality factor, $Q=\omega \tau_p$, exceeds unity ($\omega$ is the plasmon frequency). Such long relaxation times are favourably achieved in III-V heterostructures~\cite{Gusikhin_damping_reduction} and encapsulated graphene~\cite{koppens2017acoustic,bandurin2018resonant}.

The conceptual scheme of plasmon-enhanced THz detector based on 2d electron system (2DES) was put forward in Ref.~\onlinecite{Dyakonov_detection_mixing}, and realised decades later with GaAs-based heterostructures wells~\cite{Knap_resonant,Kukushkin_resonant,Peralta_resonant,Giliberti} and graphene~\cite{bandurin2018resonant}. Such detectors consist of a field-effect transistor (FET) with a 2d channel and parallel source and drain contacts. THz radiation impinging on an antenna connected between source and gate creates an alternating electric potential in the channel. Rectification by transistor nonlinearities results in a dc photovoltage between source and drain. The performance of this scheme was scarcely analysed critically, and most experimental work on transistor-based THz detectors simply copied this proposal~\cite{Knap_FET_detector_review,Roskos_THz_detector,vicarelli2012graphene}. Still, it is readily apparent that such a detector structure is almost symmetric, while structural asymmetry is the key requirement to achieve photovoltage at zero bias.

In this paper, we show that the responsivity of transistor-based THz detectors can be greatly enhanced in structures supporting highly asymmetric THz plasmon modes with singular electric fields at the contacts. Such strong fields are generally formed near keen, or thin, electrodes. Importantly, these modes are associated with oscillations of 2D electrons, and not electrons in metal contacts~\cite{Pendry_singular_plasmonics}. This makes a conceptual difference with photodetectors enhanced by plasmonic "lightning-rod" effect~\cite{nanoneedle}, allowing to push the resonance towards THz frequencies. Combination of geometric and plasmonic field enhancement results in strong nonlinear rectification, especially for nonlinearities localised near the contacts~\cite{Ryzhii_Shottky_detection,Kukushkin_Contact_rectification}. 

\begin{figure}
\includegraphics[width=\linewidth]{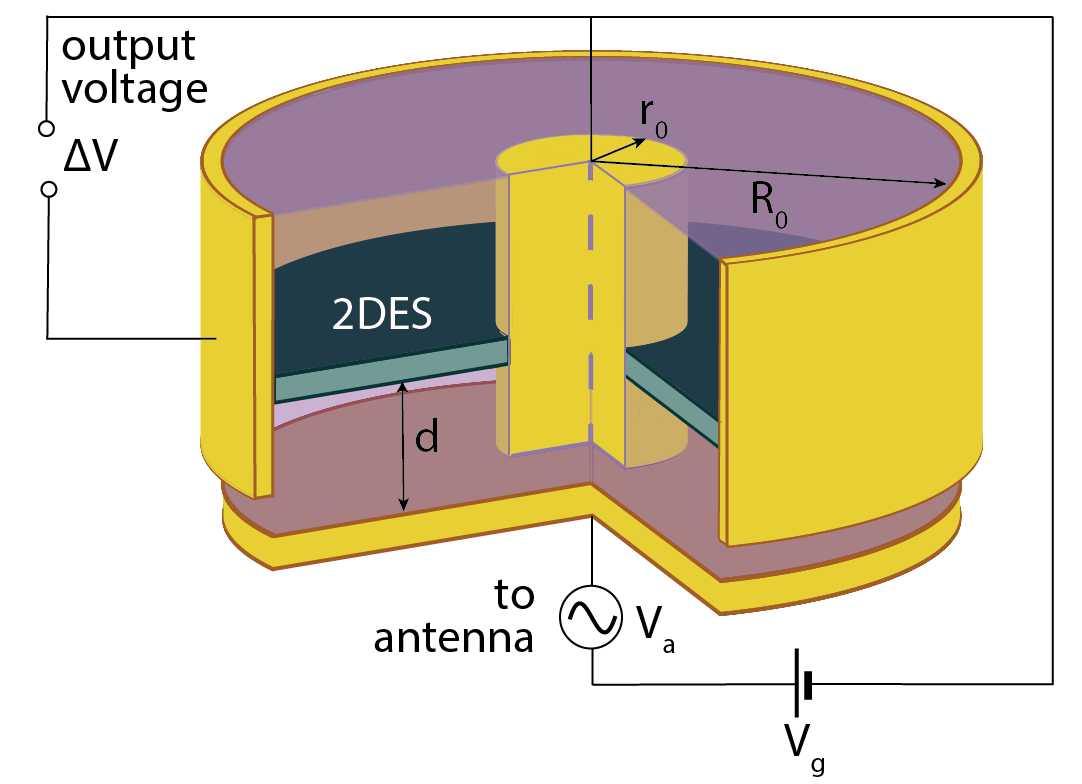}
\caption{Scheme of THz detector based on 2DES in a Corbino geometry: THz radiation from the antenna creates electric potential, $V_a$, between source and gate; dc photovoltage, $\Delta V$, due to transistor non-linearities is read out between source and drain.}
\label{fig-illustration}
\end{figure}

We consider plasmon-assisted THz rectification in a gated 2DES in the Corbino disc geometry (Fig.~\ref{fig-illustration}) as a prototypical and exactly solvable model system. We show that detector responsivity diverges as the radius of inner contact $r_0$ tends to zero according to $r_0^{-2}$, and is limited by the non-locality of rectification mechanisms. Comparing the responsivity of FET-based detectors with parallel and concentric source and drain, we find that the latter exceeds the former by an order of magnitude already for $R_0/r_0=30$.

The combined geometric-plasmonic enhancement of detector responsivity is most clearly seen via an example of a 2DES transistor in the Corbino geometry (Fig.~\ref{fig-illustration}). High-frequency radiation impinging on the antenna connected to source and gate results in an alternating voltage, $V_a\cos\omega t$. Rectification of this voltage by various nonlinearities in 2DES results in a dc photovoltage, $\Delta V$, read out between source and drain. For computational simplicity, we develop a theory of both linear response and rectification within the hydrodynamic approximation. The latter has recently been proven to hold at elevated temperatures in high-quality graphene devices~\cite{Lucas_review, bandurin2018fluidity,sulpizio2019visualizing}. Furthermore we argue that geometric and plasmonic enhancement holds for other rectification mechanisms and other transport regimes as well.

Electron motion in an axially-symmetric Corbino device is governed by Euler equation for drift velocity $u$ \begin{equation}
\label{eq-euler}
\frac{\partial u}{\partial t}+ u \frac{\partial u}{\partial r} + \frac{u}{\tau} = \frac{e}{m} \frac{\partial \varphi}{\partial r},
\end{equation}
and continuity equation for sheet density $n_s$
\begin{equation}
\label{eq-continuity}
\frac{\partial n_s}{\partial t}+ \frac{1}{r}\frac{\partial}{\partial r}\left( r u n_s \right) = 0.   
\end{equation}
In the above equations, $e$ is the electon charge, $m$ is the electron effective mass, $\tau_p$ is the momentum relaxation time, and $\varphi$ is the local electric potential in the 2DES. If gate-to-channel separation $d$ is small compared to the inner ($r_0$) and outer ($R_0$) radii of the contacts, the local gate-to-channel voltage $V_g - \varphi$ is proportional to the local carrier density   
\begin{equation}
\label{grad-ch-ap}
 C (V_g - \varphi ) = {e} n_{s},
\end{equation}
where $C = \varepsilon/4\pi d$ is the gate-channel specific capacitance, and $\varepsilon$ is the gate dielectric constant. We will show below that the above {\it local capacitance approximation} is not necessary for geometric responsivity enhancement and is considered here for analytical traceability.

The set of transport equations (\ref{eq-euler}-\ref{grad-ch-ap}) is supplemented by boundary conditions implying fixed gate-to-channel voltage at the source
\begin{equation}\label{bound1}
\varphi(r_0,t) = V_g(t) \equiv V_0 + V_a\cos(\omega t),
\end{equation}
where $V_0$ is the dc part of the gate voltage, and zero ac current at the drain
\begin{equation}
\label{bound2}  
n u_s (R_0,t) = 0.
\end{equation}
Importantly, the latter condition for {\it ac} current does not require a perfect voltmeter in the drain circuit. It is rather dictated by large inductance of bonding wires and their large reactive impedance at terahertz frequencies~\cite{Svintsov_PRApplied}. 

The linear-response ac electric potential in the channel is readily obtained from Eqs.~(\ref{eq-euler} -- \ref{bound2}):
\begin{equation}\label{phi}
\varphi_\omega(r) = - V_a\frac{Y_1(k R_0) J_0(k r) - J_1(k R_0)Y_0(k r)}{Y_0(k r_0)J_1(k R_0)-Y_1(k R_0)J_0(k r_0)},
\end{equation}
where $k = \frac{\omega}{s} \sqrt{1 + i(\omega\tau)^{-1}}$ is the plasmon wave vector, $s = \sqrt{e V_0/m}$ is the plasma wave velocity, $J_i$ and $Y_i$ are the Bessel functions of the $i$-th kind.

The obtained expression for linear-response electric potential can used as a building block to evaluate various non-linear rectification processes, e.g. due to contacts with non-linear $I(V)$-characteristics~\cite{Ryzhii_Shottky_detection} or photo-thermoelectric effects~\cite{Bandurin_dual_origin}. Here we shall focus on rectification mechanisms intrinsically present in hydrodynamic model~\cite{Dyakonov_detection_mixing}. These mechanisms include the non-linear dependence of electron fluid kinetic energy on drift velocity (hydrodynamic nonlinearity) and resistive self-mixing. Their contributions to rectified voltage are denoted as $\Delta V_{\rm hd}$ and $\Delta V_{\rm rsm}$, respectively
\begin{equation}
\label{dc v}  
\Delta V =  \Delta V_{\rm hd} + \Delta V_{\rm rsm}.
\end{equation}
Their partial contributions are given by~\cite{Sakowicz}
\begin{gather}\label{hydro v}
    \Delta V_{\rm hd} = \frac{e |E_{\omega}|^2}{4 m (\omega^2 +{\tau^{-2}}) }, \\
\label{self-mixing v}
    \Delta V_{\rm rsm} = \frac{1}{2 V_0^2} \int\limits_{r_0}^{R_0}\mathrm{Re}\left(\frac{E_\omega U_\omega^{*}}{1 - i\omega \tau} \right)dr,
\end{gather}
where $E_{\omega}(r) = -\partial\varphi_{\omega}/\partial r$ is the longitudinal electric field in the channel, and $U_\omega = V_a - \varphi_\omega$ is the local gate-to-channel voltage.

\begin{figure}[ht]
\center{\includegraphics[width=0.9\linewidth]{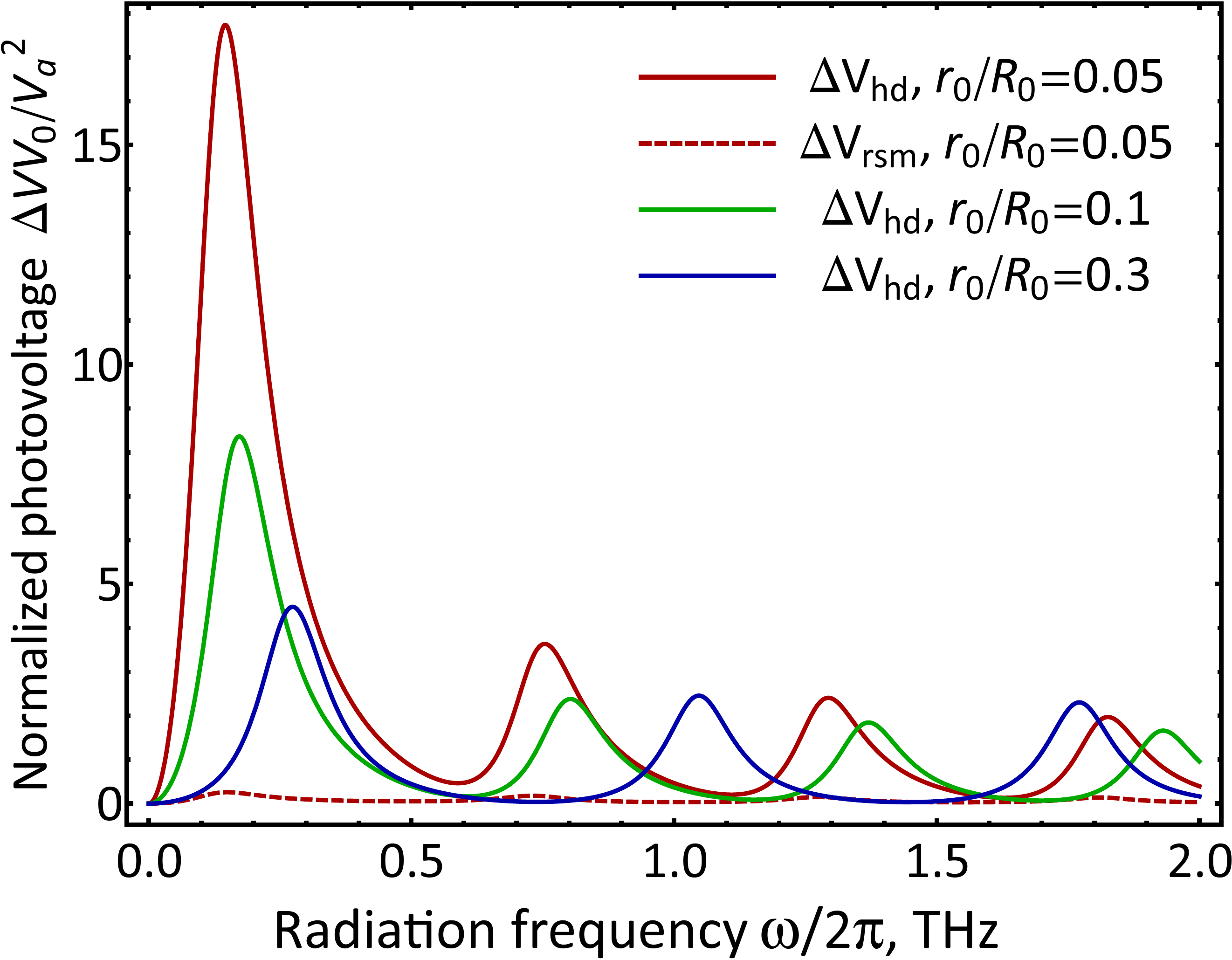}}
\caption{Normalised response of a Corbino detector vs. radiation frequency for different dimensions of inner contact, $r_0/R_0$. Contributions of hydrodynamic non-linearity, $\delta V_{\rm hd}$, and resistive self-mixing, $\delta V_{\rm rsm}$, are shown with solid and dashed lines, respectively. 2DES parameters are: $s=10^8 $ cm/s, $\tau=1$ ps, $R_0=1\,\mu$m.}
\label{fig:Responsivity1}
\end{figure}

Once the plasmon quality factor is large, $Q = \omega\tau \gg 1$, the rectified voltage is dominated by hydrodynamic nonlinearity, $\delta V_{\rm hd}/\delta V_{\rm rsm} \sim Q^2$. Indeed, the hydrodynamic nolinearity $\delta V_{\rm hd}$ is proportional to amplitude of electron velocity which, in turn, is enhanced for higher mobility. The resistive self-mixing, on the contrary, requires large ohmic resistance of the sample and is not enhanced by plasmonic effects. Numerical comparison of these contributions to rectified voltage is presented in Fig.~\ref{fig:Responsivity1} and confirms these arguments. These facts enable us to consider only the hydrodynamic contribution to rectified voltage in high-quality samples.

As the radius of the inner contact becomes smaller, the ac electric field in its vicinity is enhanced according to $E_{\omega} \approx (V_a/r_0) \ln^{-1}(e^{\gamma} k r_0/2)$, where $\gamma \approx 1.7$ is Euler's constant. At resonant frequency, the log-factor is cancelled (see Supporting Section I), and the electric field scales as $1/r_0$. This scaling coincides with the field of a field of a charged wire, which can mimic a central contact.

Geometrically-enhanced ac electric field near the inner contact leads to even higher ac velocity of electrons $u_\omega$ and further enhancement of hydrodynamic non-linearity. This result is readily seen from comparison of responsivity spectra at various ratios $r_0/R_0$ shown in Fig.~\ref{fig:Responsivity1}. The geometric enhancement is most pronounced for the lowest-frequency resonant modes; it is possible to show that rectified voltage at the first resonance is roughly $(R_0/r_0)^2$ times greater than the next one.

It is possible to covert the rectified voltage into a practical figure of detector merit, responsivity $R_V = \Delta V/P$, where $P$ is the incident electromagnetic power. The latter is related to ac voltage between source and drain via antenna radiative resistance $Z_a$, $P = V_a^2/2Z_a$. Using a typical value for dipole antennas $Z_a \approx 100$ $\Omega$, we predict the resonant responsivity of Corbino detector $R_V \approx 50$ kV/W for frequency of 0.1 THz and momentum relaxation time $\tau \approx 2$ ps. This value of responsivity exceeds by an order of magnitude that of commercially available diodes~\cite{producers}

The growing rectification capability of Corbino device with decreasing inner radius should have physical upper limits. When contact radius $r_0$ becomes comparable with gate-to-channel separation $d$, the gradual-channel-approximation becomes, strictly speaking, inapplicable. Still, it is possible to construct an exact solution for Corbino-device electrodynamics and show that (1) the gradual-channel model works fine for $r_0 < d$ (2) the growth of responsivity with decreasing $r_0$ persists for $r_0 < d$. 

The {\it exact solution} for driven electrical oscillations  in Corbino geometry is based on expansion of electric potential $\phi(r,z)$ in basis functions $\Phi_n(r,z)$ satisfying zero boundary conditions at the contacts~\cite{Svintsov_PRApplied}. The expansion is supplemented by ''contact contributions'' $V_d f_d(r,z)$ and $V_g f_g (r,z)$ representing electric potential due to gate and drain in the absence of 2DES. Therefore
\begin{equation}
    \label{eq-exact-phi}
    \varphi(r,z) = \sum\limits_{n=1}^{\infty}c_n \Phi_n(r,z) + V_d f_d(r,z) + V_g f_g(r,z).
\end{equation}
In the above expression, the gate voltage $V_g$ is fixed by antenna, the coefficients $c_n$ (strengths of $n$-th plasmon modes) are found by projection technique, and the drain voltage $V_d$ is obtained from zero-current boundary condition.

\begin{figure}[h]
\center{\includegraphics[width=0.9\linewidth]{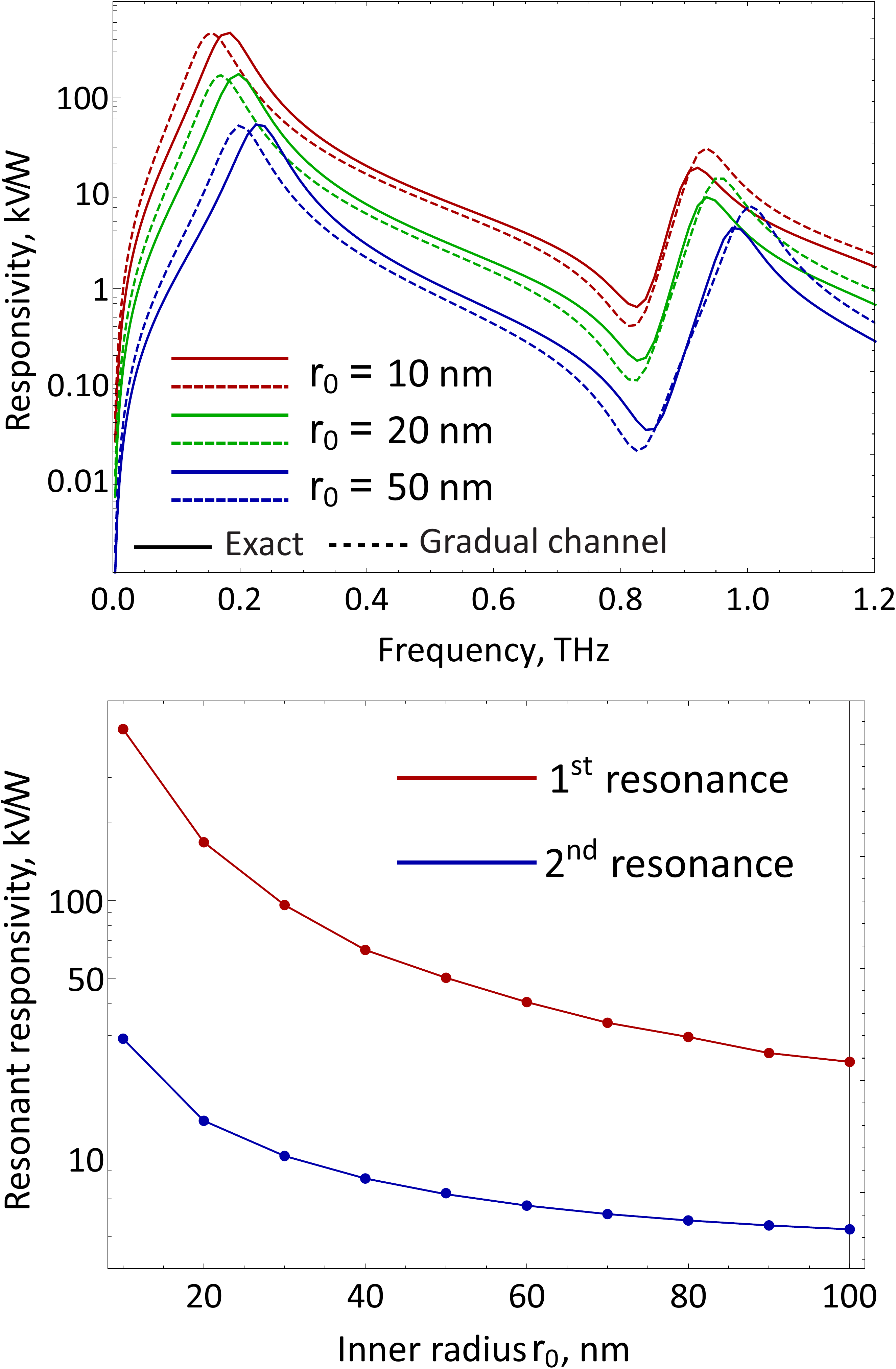}}
\caption{(a) Calculated Corbino detector responsivity vs. radiation frequency for approximate (solid line) and exact (dashed line) solutions for electric potential at different radii of the inner contact, $r_0$. (b) Responsivity at 1st and 2nd harmonics of plasmon resonance vs inner radius $r_0$. Carrier density $n_s = 10^{12} $ cm$^{-2}$, dielectric constant $\varepsilon=12.9$, momentum relaxation time $\tau_p = 2.3$ ps, $R_0=1\,\mu$m, gate-to-channel separation $d = 50$ nm, antenna resistance $Z_a = 100\,\Omega$.}
\label{fig-comparison}
\end{figure}

Explicit expressions for eigenfunctions $\Phi_n(r,z)$ can be obtained if source and drain contacts are extended to infinity in the $z$-direction. Physically, their height should much exceed the plasmon wavelength. In this model case, these functions are factorized,
\begin{gather}
    \Phi_n(x,z) = v_n(r)\psi_n(z),\\
    \psi_n(z) = e^{-|\Lambda_n z|} - e^{-|\Lambda_n (2d-z)|},\\
    v_n(r) = J_0(\Lambda_n r) - \frac{J_1(\Lambda_n R_0)}{Y_1(\Lambda_n R_0)}Y_0(\Lambda_n r),
\end{gather}
 where the eigenvalues $\Lambda_n$ are found from the zero boundary conditions $v_n(r_0) = v_n(R_0) = 0$. Once the eigen-functions are known, one readily obtains the Green's function of electrostatic problem and particular solutions $f_d$ and $f_g$ (Supporting section II).

The comparison of responsivities obtained with exact and approximate electrostatic models is shown in Fig.~\ref{fig-comparison}. One readily observes that the plate capacitor model works fine even for relatively small inner radius $r_0 = 10$ nm comparable to gate-channel separation $d = 50$ nm.

\begin{figure}[h]
\center{\includegraphics[width=0.9\linewidth]{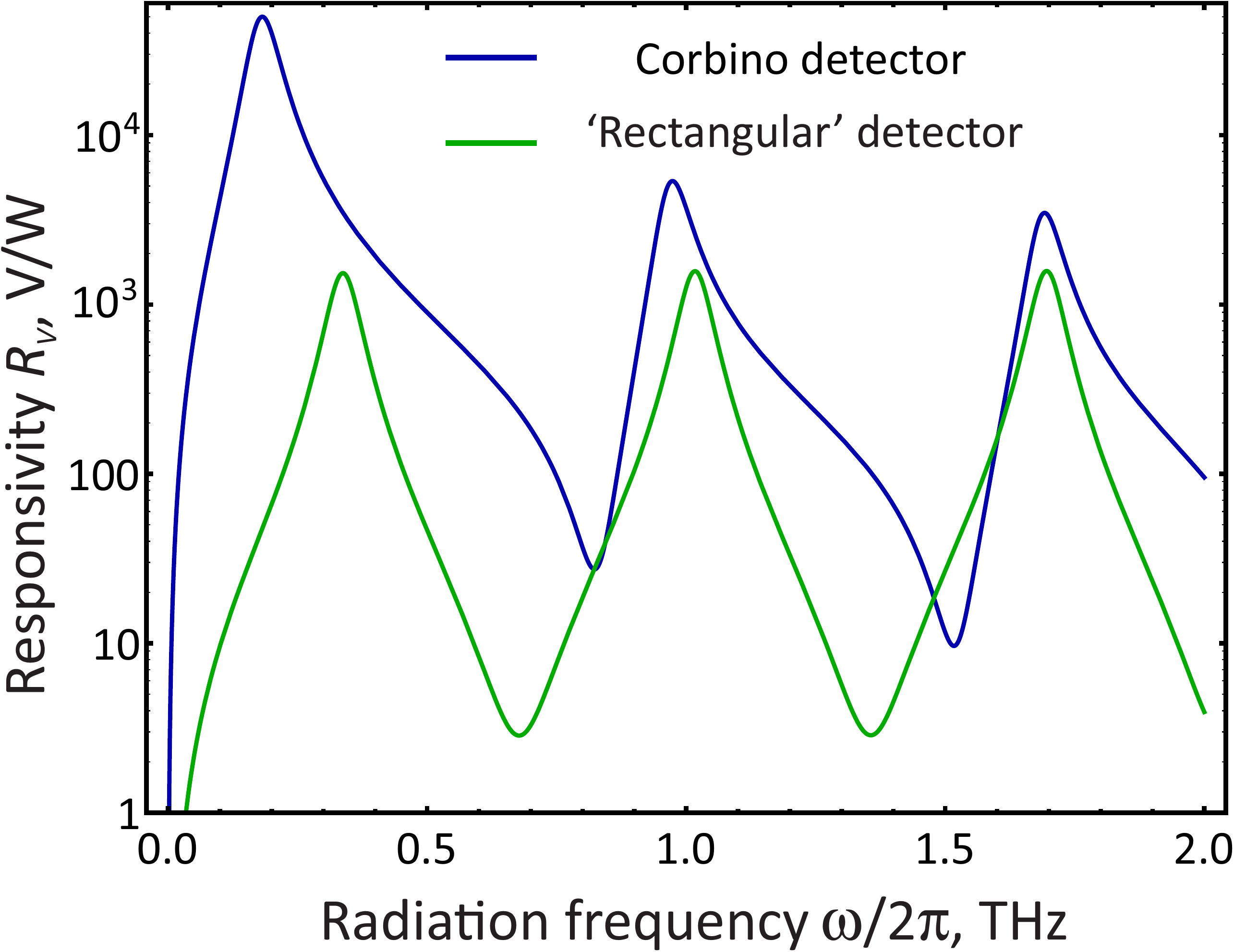}}
\caption{Responsivities of Corbino (blue line) and ''rectangular'' (green line) detectors versus radiation frequency at  $n=10^{12} $ cm$^{-2}$, $d=50$ nm, $\varepsilon=12.9$, $\tau=2.3$ ps, $R_0=1\,\mu$m, $r_0=0.03 R_0$, antenna resistance $Z_a = 100\,\Omega$. }
\label{fig:Corbino_vs_DS}
\end{figure}

Now we turn to  discuss the physical limitations of our model. The particular value of external responsivity would depend on coupling between antenna and 2DES; the assumed relation between incident power and gate-drain voltage $V_a^2 =2 P Z_a$ is valid if antenna acts as a perfect voltage source. Physically, it is valid if antenna impedance is less than input impedance of transmission line formed by gate plate and kinetic inductance~\cite{bandurin2018resonant,Aizin_TL_model}. To show the prospect of proposed detector being independent of coupling details, we compare the responsivity of Corbino detector and original proposal of Ref.~\onlinecite{Dyakonov_detection_mixing} (''rectangular'' FET with parallel source and drain contacts) calculated under identical conditions (Fig.~\ref{fig:Corbino_vs_DS}). We readily observe that responsivity of Corbino device exceeds that of ''rectangular'' FET by more than an order of magnitude, especially in the vicinity of the lowest resonance. As the responsivity of original proposal has been measured in numerous works and was shown to exceed 1 kV/W~\cite{bandurin2018resonant,Lisauskas_imaging}, the estimate for Corbino device exceeding 10 kV/W looks realistic.

The enhanced responsivity of Corbino detector should hold for other local rectification mechanisms different from hydrodynamic nonlinearity. Particularly, contacts between metal and 2DES can display rectifying behaviour either due to non-linearity of the Schottky junction $I(V)$-characteristic~\cite{Ryzhii_Shottky_detection}, or due to photo-thermoelectric effect at the metal-2DES junction~\cite{Nikitin_fast_and_sensitive,Bandurin_dual_origin}. Moreover, detectors made of graphene were predicted to exhibit stronger photoresponse owing to graphene's linear band dispersion~\cite{Tomadin1,Tomadin2}. Last but not least, it was recently shown~\cite{Pseudo_Euler} that ''hydrodynamic nonlinearity'' persists in the ballistic regime as its underlying origin is the non-linear dependence of electron energy on velocity. Importantly, in any of these cases, the non-linear response is approximately proportional to squared ac electric field at the contact $|E_\omega(r_0)|^2$, as in the considered case of hydrodynamic nonlinearity. As the field enhancement by plasmonic and geometric effects is independent of assumed rectification physics, high responsivity of Corbino devices should retain for other rectification mechanisms. We also note that an inverse process of dc current conversion into plasmons (plasma instability~\cite{DS_instability,Lucas_DS,LucasPolini}) is also favoured by large electric fields of plasmon modes at the contacts~\cite{Pert_theory}, and the instability growth rate should also scale as $r_0^{-2}$. This remarkable scaling needs to be included in the analysis of current-driven plasmon instability in the Corbino geometry~\cite{Sydoruk_instability}.

Growth of responsivity with decreasing $r_0$ should be limited by non-locality of rectification. The hydrodynamic non-linearity is expected to saturate at $r_0 \sim l_{e-e}$, where $l_{e-e}$ is free path due to electron-electron collisions; the Schottky junction nonlinearity would saturate at $r_0 \sim l_{dep}$, where $l_{dep}$ is the depletion length; the photo-thermoelectric effect would saturate at $r_0 \sim l_{T}$, where $l_T$ is the heat diffusion length. Last but not least, the decrease in inner contact radius would increase the total resistance of the device $R_{tot}$. However, the latter scales with $r_0$ only logarithmically, $R_{tot}\propto \ln R_0/r_0$, and should not have much effect on current resposnivity $R_I = R_V/R_{tot}$ and noise-equivalent power ${\rm NEP} \propto \sqrt{R_{tot}}/R_V$.

To sum up, we have shown a strong enhancement of responsivity in THz and sub-THz detectors based on 2D electron systems in Corbino geometry. The enhancement appears due to a combination of plasmon resonance and electric field enhancement near the central contact of small radius. We have shown the possibility to achieve external responsivity exceeding 10 kV/W at 200 GHz frequency at realistic sample dimensions and momentum relaxation times, if the photoresponse is governed by hydrodynamic nonlinearity. The proposed effect of combined geometric-plasmonic responsivity enhancement should be observable not only in Corbino geometry with small central contact, but also in other structures with keen or sharp metal electrodes and can be extended to other detection mechanisms, e.g. of photo-thermoelectric origin.
\newline{}

This work was supported by grant 18-72-00234 of the Russian Scientific Foundation. The authors thank D. Svintsov, A. Tomadin, V. Muraviev, W. Knap and A. Lisauskas for helpful discussions, and Yu. Kharchenko for designer's perspective.
\newline{}

The data that support the findings of this study are available from the corresponding author upon reasonable request.
\newline{}

See the supplementary material for the calculation of resonant responsivity and exact solution for driven oscillations in Corbino geometry. 
\newline{}

\section{Supporting information}
\subsection{Resonant responsivity}
We are able to analytically maximize the hydrodynamic voltage in high-Q regime for resonant frequencies $\omega_n$ determined by the zeros denominator of Eq. (\ref{phi}):

\begin{equation}\label{hydro1}
	\Delta V_{hd}=V_a^2\frac{\tau^2}{R_0^2s^2} \frac{D_1^2(r_0,R_0)}{\left(\frac{r_0}{R_0}D_1(r_0,R_0) + D_0(r_0,R_0)\right)^2},
\end{equation}
where
$D_{i}(r_0,R_0)= \begin{vmatrix}
Y_{i}(\frac{\omega_{r}}{s} r_0)&J_{i}(\frac{\omega_{r}}{s} r_0)\\
Y_{i}(\frac{\omega_{r}}{s} R_0)&  J_{i}(\frac{\omega_{r}}{s} R_0)
\end{vmatrix}$, $i=1,2.$
At small $r_0$ we have $D_{1}(r_0,R_0) \gg D_{0}(r_0,R_0)$ because $D_{1}(r_0,R_0) \simeq J_{1}(\frac{\omega_{r}}{s} R_0) Y_{1}(\frac{\omega_{r}}{s} r_0) \simeq \frac{const}{r_0}$ and $D_0(r_0,R_0) \simeq J_{0}(\frac{\omega_{r}}{s} R_0) Y_{0}(\frac{\omega_{r}}{s} r_0) \simeq const \, \mathrm{ln}(\frac{\omega_{r}}{s} r_0)$. At these conditions, Eq. (\ref{hydro1}) can be simplified:
\begin{equation}\label{hydro_small_r0}
	\Delta V_{hd}  \propto V_a^2\frac{\tau^2}{r_0^2s^2}
\end{equation}

\subsection{Exact solution for driven oscillations in Corbino geometry}

The central object to find electric potential due to gate and drain in the absence of 2DES is the Green's function of electrostatic problem. Once the eigenmodes of Laplace operator in Corbino geometry $\Phi_n(r,z)$ are known, the Green's function is expressed as:
\begin{multline}
	\label{eq-Green-exact}
	G(r,r',z,z') = \\
	\frac{1}{4\pi}\sum\limits_{n=1}^\infty \frac{v_n(r')v_n(r)}{\Lambda_n}\left(e^{-\Lambda_n|z-z'|} - e^{-\Lambda_n|z+z'-2d|} \right),
\end{multline}

where
\begin{equation}
	v_n(r) = J_0(\Lambda_n r) - \frac{J_1(\Lambda_n R_0)}{Y_1(\Lambda_n R_0)}Y_0(\Lambda_n r),
\end{equation}
and $\Lambda_n$ are found from the condition $v_n(r_0) = 0$. We do not consider the modes with angle-dependent electric potential $\propto e^{i m \theta}$, as they do not couple to radially-symmetric field.

The functions $f_d(r,z)$ and $f_g(r,z)$ in Eq. (10) are electric potentials created in all space when the potential of drain (gate) is set to unity. These both can be found using the fundamental solution of the Laplace equation:
\begin{equation}
	f_d(r,z) = -R_0\sum\limits_{n=1}^\infty\frac{v_n'(R_0)v_n(r)}{\Lambda_n^2}(1-e^{-\Lambda_n(d-z)});
\end{equation}
\begin{equation}
	f_g(r,z) = \sum\limits_{n=1}^\infty v_n(r)e^{-\Lambda_n(d-z)}\int\limits_{r_0}^{R_0}v_n(r')r'\,\mathrm dr',
\end{equation}
$v_n'(r)$ denotes the derivative of $v_n(r)$.

\bibliography{dialogue}

\end{document}